# Shape Memory Polymer Resonators as Highly Sensitive Uncooled Infrared Detectors

Ulas Adiyan[1], Tom Larsen[2], Juan José Zárate[1], Luis Guillermo Villanueva[2], Herbert Shea[1]

**Abstract**

Uncooled InfraRed (IR) detectors have enabled the rapid growth of thermal imaging applications. These detectors are predominantly bolometers, where the heating of pixel from incoming IR radiation is read out as a resistance change. Another uncooled sensing method is to transduce the IR radiation into the frequency shift of a mechanical resonator. We present here a highly sensitive, simple to fabricate resonant IR sensor, based on thermo-responsive Shape Memory Polymers (SMPs). By exploiting the phase-change polymer as the transduction mechanism, our approach provides 2 orders of magnitude improvement of the temperature coefficient of frequency (TCF). The SMP has very good absorption in IR wavelengths, obviating the need for an absorber layer. A Noise Equivalent Temperature Difference (NETD) of 22 mK in vacuum and 112 mK in air are obtained using f/2 optics. Such high performance in air eliminates the need for vacuum packaging, paving a path towards flexible IR sensors.

[1] Soft Transducers Laboratory, École Polytechnique Fédérale de Lausanne (EPFL), 2002 Neuchâtel, Switzerland. [2] Advanced NEMS Group, École Polytechnique Fédérale de Lausanne (EPFL), 1015 Lausanne, Switzerland. Correspondence and requests for materials should be addressed to H.S. (email: herbert.shea@epfl.ch)

Thanks to progress in the infrared (IR) detection, IR imaging is rapidly being adopted in a very broad range of fields such as thermography, medical diagnostics, firefighting, autonomous driving, food inspection and security[1–6]. The two major IR detection technologies are photon detectors (typically cooled) and thermal detectors (generally uncooled). Photon detectors rely on the generation of electron/hole pairs within the detector material when exposed to IR radiation. They are the fastest and the most sensitive IR detection technology; however, they require operation at cryogenic temperatures to minimize thermally generated electron-hole pairs. Cooling adds significantly to device size and cost. Thermal detectors transduce IR radiation by measuring a temperature-dependent physical property such as electrical resistance in bolometers[7], electrical polarization in pyroelectric detectors[8], electrical voltage in thermopiles[9], and displacement in thermomechanical detectors[10]. As room temperature detectors, they do not require active cooling, hence can be smaller and more power efficient[11]. However, they lag behind photon detectors in terms of sensitivity and response time. Resonant IR sensors[12–19], whose mechanical resonant frequency have thermal dependence, can be a breakthrough for thermal detectors by providing extremely high sensitivity owing to the low noise associated with frequency measurements[20].

Noise Equivalent Temperature Difference (NETD) is a key figure of merit for IR sensors in order to quantify their sensitivity. It refers to the minimum distinguishable temperature change at the radiation source (e.g. IR target). Equation (1) shows the generalized analytical expression[10] of the NETD. $\Delta T_T$ is the temperature change at the radiation source, namely the target temperature change, and SNR is the signal to noise ratio of the infrared detector output, where $V_s$ is the signal level (i.e. signal voltage) and $V_N$ is the total noise (i.e. RMS noise voltage) within the system bandwidth[21].

$$NETD = \frac{\Delta T_T}{SNR} = \frac{\Delta T_T}{V_s/V_N} \tag{1}$$

Equation (2) shows the modified analytical expression for the NETD of the resonant IR sensors[14] using Equation (1).

$$NETD = \frac{\Delta T_T}{P_{inc}} \frac{P_{inc}}{P} \frac{P}{\Delta T_D} \frac{\sigma_A}{TCF} \tag{2}$$

$P_{inc}$ is the incident power on the detector, $P$ is the absorbed power by the detector, $\Delta T_D$ is the temperature change at the detector, $\sigma_A$ is the Allan deviation[22] (a measure for the noise in the frequency), and TCF is the temperature coefficient of the frequency shift. To enhance sensitivity, one wishes to minimize each factor of Equation (2). We first have $(\frac{\Delta T_T}{P_{inc}})$, which is related to the IR optical system and it falls outside of the scope of this paper; then $(\frac{P_{inc}}{P})$, which is the inverse of the absorbance; and then we have $(\frac{P}{\Delta T_D})$, which is the thermal conductance of the sensor (also including conduction to the air if not vacuum-packaged). Maximizing the absorbance[23] and minimizing the thermal losses[17,24] of the sensing area improve the sensitivity. Finally, the last term $(\frac{\sigma_A}{TCF})$ consists of the two key parameters that are specific for resonant IR sensors. High sensitivity can be achieved by increasing the TCF[25,26] and/or improving the frequency stability[27–29].

In this work, we report a highly sensitive, simple to fabricate and low-cost resonant IR sensor based on the strong temperature dependence of the Young's modulus of thermo-responsive Shape Memory Polymers (SMPs)[30–32]. SMPs are programmable phase-change materials that can memorize a permanent shape, be deformed and fixed in a temporary shape under given conditions, and later, upon external command, relax to their original permanent shape[33–38]. We present the first use of the phase transition as a transduction mechanism for IR detection. Our SMP resonators provide unprecedented responsivity owing to the highest TCF ever reported as an IR sensor. The SMP material itself not only provides the transduction mechanism but is also a good absorber in the long-wave IR (LWIR) range (i.e., from 7 to 14 µm), avoiding the need for an additional absorber layer. The SMP material has low thermal conductivity, enabling good thermal isolation from the environment. These characteristics of the SMP material enable high sensitive IR detection not only in vacuum but also at atmospheric pressure.

## Results

**Working Principle.** Figure 1 summarizes the working principle of the resonant IR detection technology, based on SMP resonators.

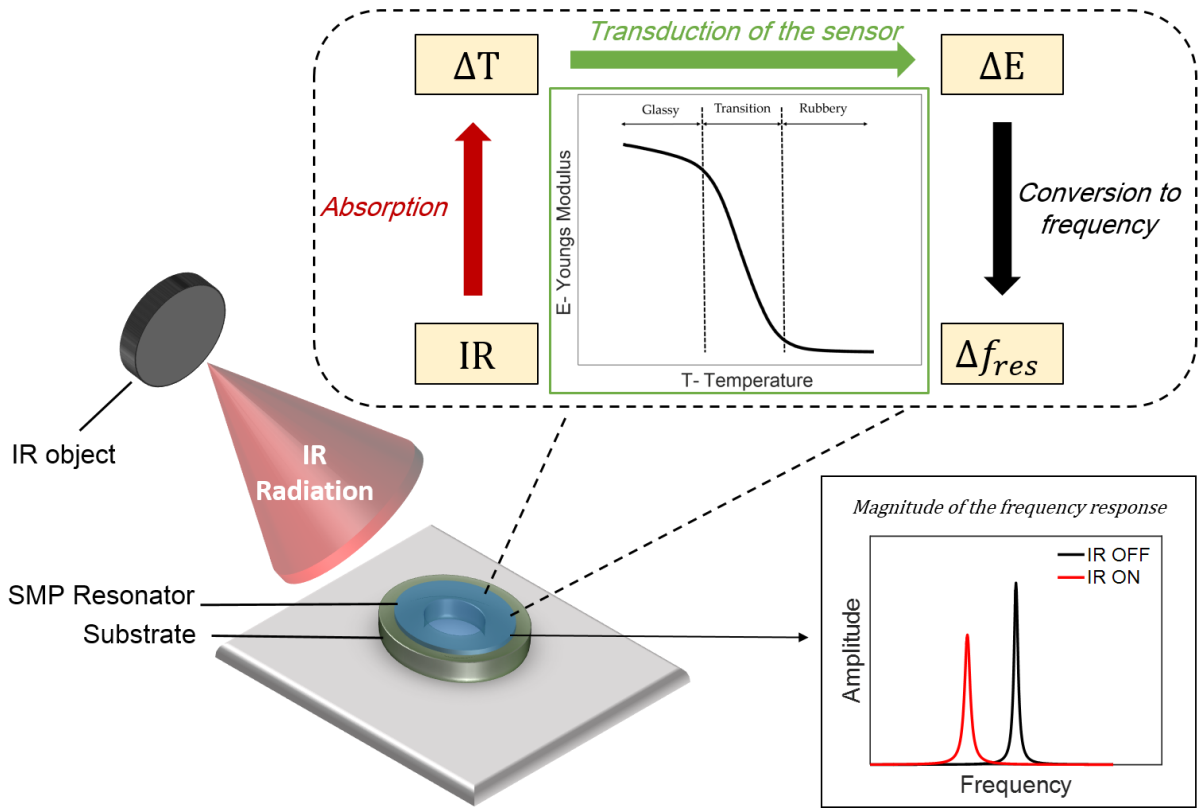

**Figure 1:** Overview for the resonant IR SMP resonant sensor and the working principle. The incident IR radiation from the IR object causes a temperature change on the SMP resonator. This leads to a change in the Young's modulus of the SMP material, which shifts the resonance frequency of the SMP resonator. Thus, the change in the IR object's temperature can be detected by a frequency readout scheme[39].

The incident IR radiation from an object induces a temperature change on the SMP resonator, depending on IR optics, the absorbance and the thermal isolation of the sensor. This temperature change leads to a change in the Young's modulus of the SMP material, shifting the resonance frequency of the resonator. The shift in the resonance frequency can be detected using a frequency readout scheme[39], and can be converted to a temperature change by calibrated IR sources[6].

The magnitude of the change in the resonance frequency is determined by the TCF of the sensor. TCF is the relative shift in resonance frequency due to sensor temperature change, which can be written as $TCF = \frac{1}{f_{Res}} \frac{\partial f_{Res}}{\partial T_D}$ [1/K], where $f_{Res}$ is the resonance frequency. The larger the TCF, the larger the change of the resonance frequency for a given temperature change. For our sensor, the rate in the change of the Young's modulus varies with the temperature of SMP, so does the TCF. In principle, the maximum TCF occurs near the glass transition temperature ($T_{glass}$) of the SMP material used and our purpose is to operate the sensors around $T_{glass}$ to achieve the highest TCF possible.

**SMP Material Characterization.** Supplementary Figure 1 shows the dynamic mechanical analysis (DMA) measurement for the commercial MM4520 SMP material, which has a $T_{glass}$ of 45 °C. The Young's modulus ($E$) is plotted vs. temperature from 20°C to 80°C. $E$ drops from around 1700 MPa to 10 MPa in this range. This corresponds to a factor of 13 change in the resonance frequency of a resonator made out of this material within this 60 °C span. We plot the thermal coefficient of the Young's modulus $TCE = \frac{1}{E} \frac{\partial E}{\partial T_D}$, which has a maximum value of around $TCE_{peak} = 0.2$ K$^{-1}$, and is located in the glass transition region at around 50 °C. This $TCE_{peak}$ value implies a $TCF_{peak}$ of 0.1 K$^{-1}$ (i.e., 10%). Having an extremely high TCF may result in a very sensitive resonant IR sensor, if the resonance frequency can be tracked accurately, which is linked to the Q-factor of the resonator. The Q-factor is related to the energy loss of the system, so that, an intrinsic quality factor ($Q_E$) can be defined using the inverse of the loss factor of the material i.e. $Q_E = |tan(\delta)|^{-1} = E'/E''$. $E'$ is the storage modulus and $E''$ is the loss modulus that can be obtained from the DMA measurements. $Q_E$ is plotted vs. temperature in Supplementary Figure 1. $Q_E$ drops substantially around the glass transition temperature. There is a trade-off between the Q-factor and the TCE or TCF, since the maximum loss for the SMP material coincides with the $TCE_{peak}$ due to the high viscoelasticity around $T_{glass}$. Operating the IR SMP resonator away from $T_{glass}$ results in higher Q-factor, but lower TCF. As a consequence, the optimum operation temperature must be determined for each particular resonator and might very well be away from $T_{glass}$.

***SMP Resonator Characterization.*** We carried out tests with a circular SMP resonator, which has a radius of 520 µm and a thickness of 10 µm. The resonator is placed onto a Polymethyl methacrylate (PMMA) substrate having a thickness of 1 mm. The fabrication and the assembly of the SMP resonators are explained in the Methods section.

The SMP resonator was attached onto a piezo-disk actuator and placed on a heater with a PID temperature controller system, which controls the operation temperature of the resonator ($T_{sub}$), using a thermistor placed on the substrate (Supplementary Figure 2). The temperature was increased from $T_{sub}$=25 °C to $T_{sub}$=50 °C in steps of 5 °C. Firstly, the change in resonance frequency and Q-factor were monitored by measuring the displacement vs. frequency using a Laser Doppler Vibrometer (LDV). Then, TCF and the frequency stability noise ($\sigma_A$) of this resonator were determined for every operation temperature from 25 to 50ºC. Lastly, characterization tests for IR detection to determine experimentally the NETD and response time were carried out with the same resonator.

**Resonance frequency and Q.** Figure 2.a shows the resonance frequency of the 1st flexural mode with respect to temperature in vacuum (~ $10^{-3}$ Pa). For each $T_{sub}$, the measured resonant responses of the SMP membrane are fitted to a Lorentzian function to extract the resonance frequencies and the Q-factors. Supplementary Figure 3 shows the frequency response of the SMP membrane for 2 different operation temperatures ($T_{sub}$=25 °C – glassy state, $T_{sub}$=45 °C – transition state). The resonance frequency vs temperature measurements were compared with analytical values computed using the Young's modulus ($E$) vs. temperature data from the DMA measurements (Supplementary Figure 1), by applying the expression for the natural frequencies of a circular membrane from ref. [39]. The experimental data agrees well with the analytical solution (Figure 2.a), with the change in the resonance frequency scaling as $f_{res} \sim \sqrt{E(T)}$).

Figure 2.b plots the Q-factor of the membrane in vacuum vs. temperature ($T_{sub}$). Both the measured Q and the intrinsic material quality factor $Q_E$ from the DMA results agree quite well. Q has the highest value near room temperature (glassy state, Q ≈ 43), it drops down to Q ≈ 1-2 in the transition region. While there are many possible energy dissipation factors such as air damping and anchor losses[40], the dominant energy loss mechanism for our case is the intrinsic material loss.

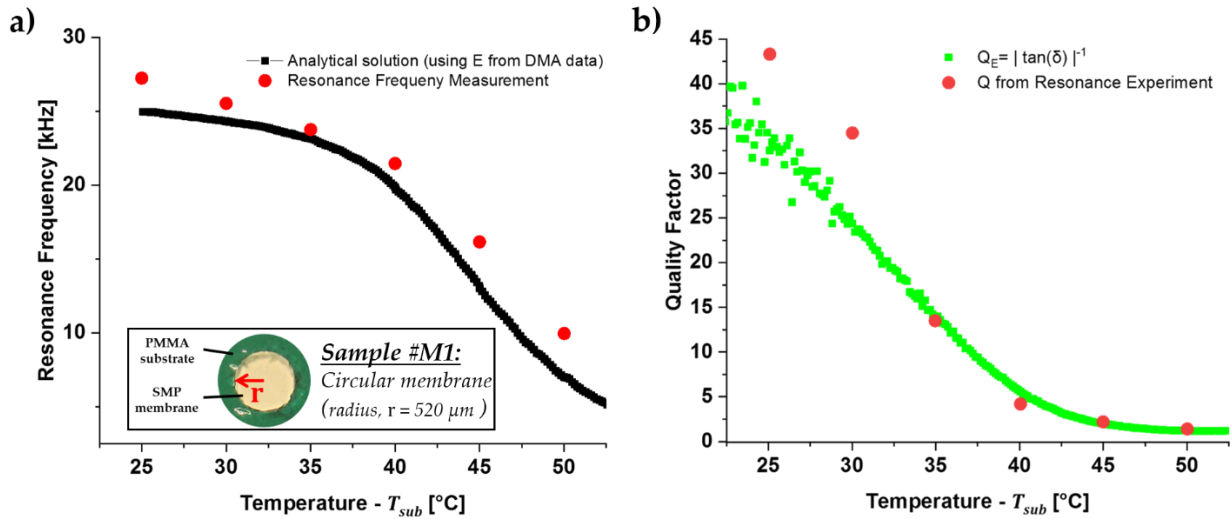

**Figure 2:** Resonance frequency and quality factor of a SMP resonator vs. temperature in vacuum. Both plots show measured data and the analytical solution based on DMA data.

***TCF, frequency stability and ultimate sensitivity.*** After measuring the resonance frequency of the SMP resonator as a function of temperature, it is straightforward to calculate the TCF: $TCF = \frac{1}{f_{Res}}\frac{\Delta f_{Res}}{\Delta T_{sub}}\ [1/K]$. By using the TCF and frequency stability noise, which is characterized by the Allan deviation (AD) - $\sigma_A$, one can calculate the minimum detectable change in the substrate temperature: $\delta T_{sub} = \frac{\sigma_A}{TCF}$. To this aim, we investigate the AD and TCF for different $T_{sub}$.

Figure 3.*a.* shows the (best-measured) AD measurements of the SMP resonator as a function of integration time (τ) for operation temperatures between 25 - 50 °C. The minimum AD values are measured around 25 - 30 °C, which is 2 order of magnitude higher than typical MEMS resonators from the literature having similar masses[22]. As we increased the operation temperature, the frequency stability degrades. This is principally due to the decrease in the Q-factor of the resonator, as the AD is inversely related to the Q-factor of the resonator[22]. The amplitude of the output signal (resonator displacement) is another important parameter that improves the frequency stability[41]. Since our detection setup was limited to the maximum optical signal, we operated the resonator in all cases close to that limit. Therefore, as we increased the operation temperature, we increased the piezo driving voltage to compensate for the reduction in Q.

In terms of integration time, the maximum frequency stability occurs between τ = 200 ms and τ = 400 ms depending on the operation temperature. The frequency stability degrades above this interval, most probably due to thermal drift. For the temperature detection sensitivity analysis, we used the AD values corresponding to an integration time of τ=500 ms, as this matches the sensor thermal time constant. (See Supplementary Note 1 for scaling down to 52 µm resonator that achieves video rate applications with a computed time constant of 7 ms).

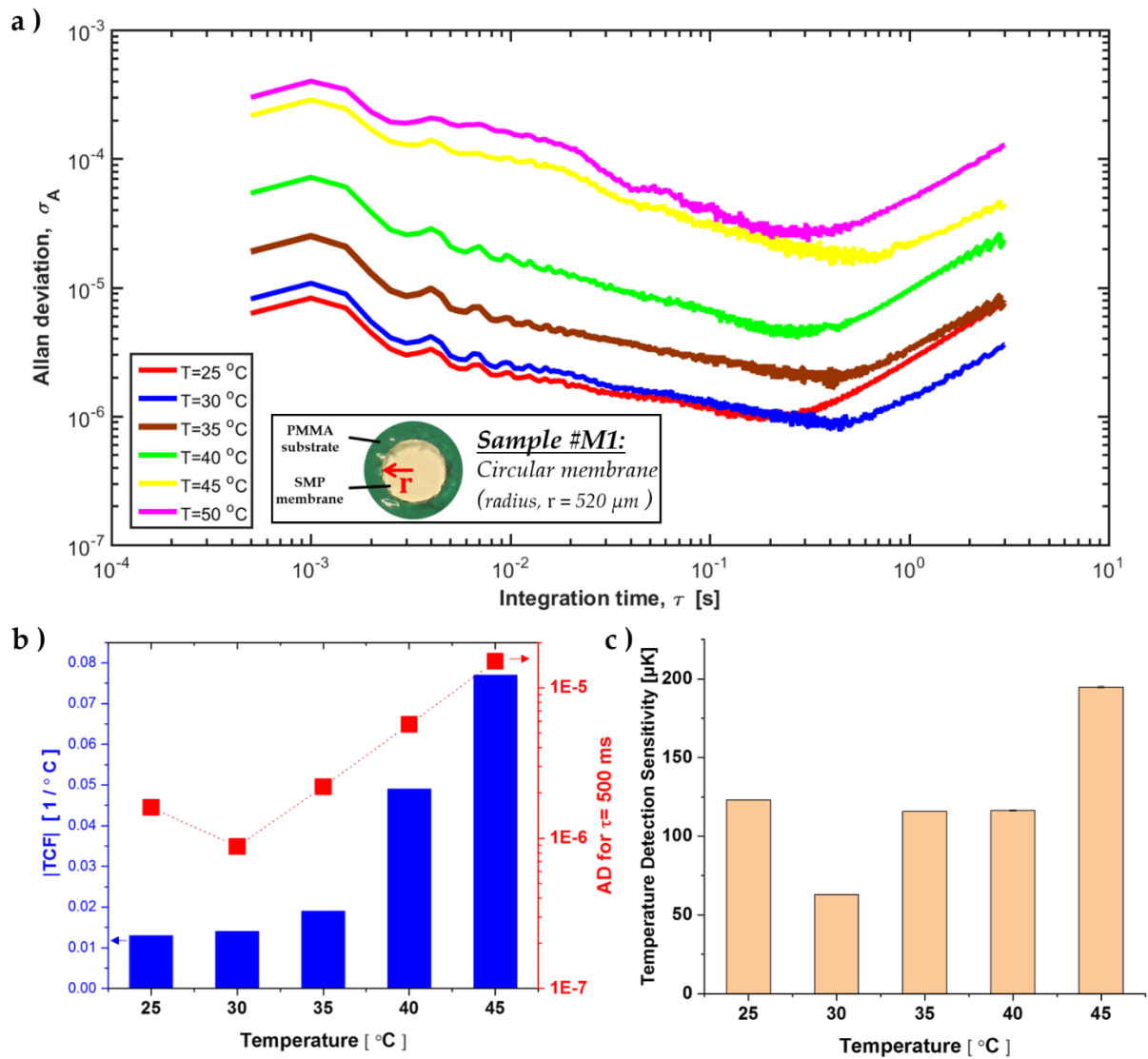

Figure 3: *a-) Allan Deviation measurements as a function of integration time for different operation temperatures for a circular SMP membrane in vacuum (~ $10^{-3}$ Pa). Please see Methods for the details of the AD measurements. b-) TCF (left y-axis) and AD measurement (right y-axis) for an integration time of τ=500 ms vs. operation temperature ($T_{sub}$). c-) Computed temperature detection sensitivity ($\delta T_{sub} = \frac{\sigma_A}{TCF}$) vs. operation temperatures. The best derived sensitivity is* 63 µK *at* $T_{sub}$=*30 °C operation temperature.*

Figure 3.b shows the measured TCF and Allan Deviation values for an integration time of τ=500 ms. The TCF were calculated according to the normalized frequency change within each 5 °C step using the measurements from Figure 2.a. The best TCF is around 8% at the operation temperature of $T_{sub}$=45 °C; which is 2 order of magnitude higher than the best TCF reported so far for IR resonant sensors[25]. Since there is a trade-off between the TCF and Q, the operation temperature of $T_{sub}$=45 °C has very high AD value for the integration time τ=500 ms as it is illustrated in Figure 3.b. Even though the transition region provides higher TCF, the Q-factors are very low, which eventually degrades the frequency stability and the sensitivity.

For each data point in Figure 3.b, the minimum detectable temperature change on the substrate ($\delta T_{sub}$), which corresponds direct temperature detection sensitivity, is calculated. The best temperature detection sensitivity is 63 µK at $T_{sub}$=30 °C operation temperature, as shown in Figure 3.c. This value implies that the simple SMP resonant sensor architecture offers temperature detection sensitivity comparable or better than the performances of the state-of-the-art photonic temperature sensors with far more complex structures[42,43].

***NETD measurements.*** The experimental setup for the characterization of the IR SMP sensors is illustrated in Supplementary Figure 4. The IR radiation from an object is transmitted and focused onto the SMP resonator. The absorbed IR radiation creates a temperature change on the IR sensor, $\Delta T_D$, which is only a small fraction of the target temperature change, $\Delta T_T$. $\frac{\Delta T_D}{\Delta T_T}$ represents the detector to target temperature ratio which depends on the IR optics, the area, the absorbance of the IR sensor and the thermal conductance (See Supplementary Note 2 for detailed analysis).

The absorbance of the 10 µm thick SMP sample was measured using Fourier transform infrared (FTIR) spectrometer throughout the LWIR spectrum (Supplementary Figure 5). The 10 µm thick SMP film absorbs 44% of the IR radiation in the 7-14 µm spectral range (See Methods for the absorbance measurements). Since the SMP material is a good absorber in LWIR range, there is no need for additional absorber layer. To test bare SMP resonators as IR sensors, the IR radiation emitted from a black resistive heater was modulated using a rotating chopper and focused on the sensor through a lens system (see Methods for the details). For the IR detection tests described in this section, we used the same resonator characterized before, which has a thickness of 10 µm.

Figure 4.a shows the IR response of the circular resonator, for two different temperature at the IR source. This leads to two different temperature differences of the source with respect to the ambient, which is referred as the blackbody temperature difference ($\Delta T_{bb}$). The temperature for the resonator substrate was maintained at $T_{sub}$=30 °C (since it is the optimal operation temperature for the device) and the resonator was operated in vacuum (~ $10^{-3}$ Pa). The temperature changes of $\Delta T_{bb}$ = 5 °C and $\Delta T_{bb}$ = 15 °C at the IR source result in a frequency shift of $\Delta f_1 \approx$ 5 Hz and $\Delta f_2 \approx$ 15 Hz respectively. The frequency tracking method is explained in the Methods. To find the NETD of the IR sensor, we used the following expression: $NETD = \sigma_A / TCF_{bb}$ where the $TCF_{bb}$ is the TCF of our sensor with respect to the 1 °C blackbody temperature difference and can be expressed as $TCF_{bb} = \frac{1}{f_{res}} \frac{\partial f_{res}}{\partial T_{bb}}$.

We obtained the *NETD* as low as 22 mK. Although a moving average filter was used to smooth the modulated IR response, this has no effect on the NETD calculation, because only AD measurements were used for the noise calculations. In addition, the thermal time constant was also deduced as $376 \pm 8$ ms based on the modulated IR response (Supplementary Figure 6), which is an important figure of merit to evaluate the response time and the potential of the sensor for real time video applications (see Supplementary Note 1).

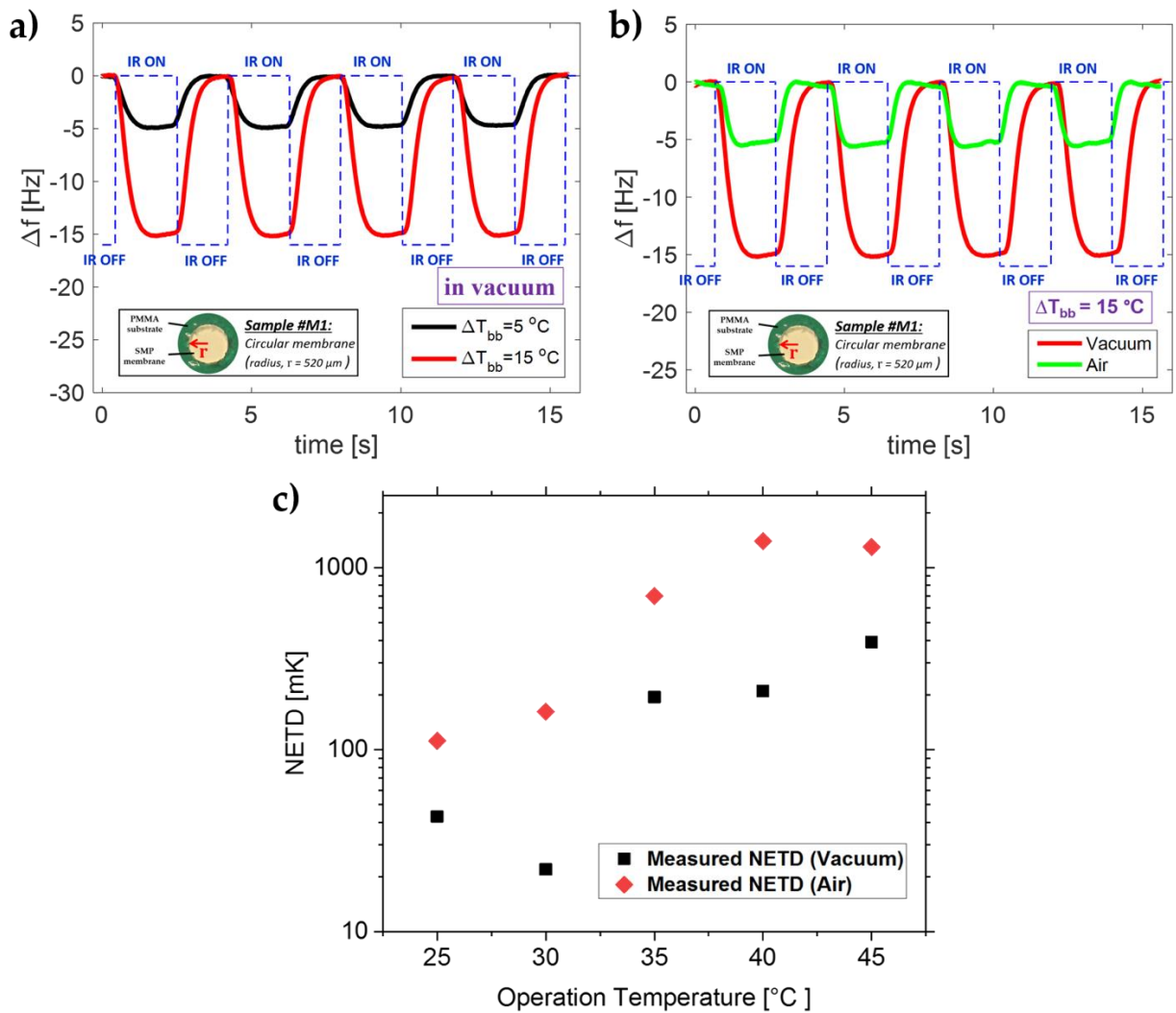

Figure 4: *a-) The frequency response of the IR sensor in vacuum at $T_{sub}$=30 °C operation temperature when the target temperature is periodically changed by $\Delta T_{bb}$=5° and by $\Delta T_{bb}$*

*=15°. A moving average filter (N=1001) is used to smooth the data, which was acquired with an acquisition rate of 25 kHz. b-) The frequency response of the IR resonator at $T_{sub}$=30 °C operation temperature in vacuum and at 1 atmosphere. A moving average filter (N=1001) is used to smooth the data, which was acquired with an acquisition rate of 25 kHz. c-) Measured NETD vs. operation temperature in vacuum and in air.*

Figure 4.b shows the IR response of the resonator, in vacuum (~ $10^{-3}$ Pa) as well as in atmospheric pressure at the operation temperature of $T_{sub}$=30 °C. A temperature change of $\Delta T_{bb}$ = 15 °C at the object results in a frequency change of $\Delta f_{air}$ ≈ 5.4 Hz in air as compared to $\Delta f_{vacuum}$ ≈ 15 Hz in vacuum. The thermal conductance from the membrane to the substrate through the air degrades the detector to target temperature ratio (See Supplementary Note 2), causing a decrease in the signal level. This effect can be seen in Figure 4.b, where the thermal time constant of the membrane in air is lower than in vacuum. In order to determine the NETD in air, similar AD measurements were performed at atmospheric pressure. The NETD at $T_{sub}$= 30 °C in air pressure was calculated as 162 mK. Even though NETD in air is almost 7 times worse than in vacuum, it is still comparable to the NETD from the state-of-the-art resonant IR sensors that operate in vacuum [14,44].

Lastly we measured the NETD for operation temperatures between 25 - 45 °C in vacuum and air (Figure 4.c). At 25 and at 30 °C, the NETD is less than 45 mK (best NETD of 22 mK) in vacuum and 160 mK (best NETD of 112 mK) in air. We compared the measured NETDs with the predicted values in Supplementary Note 3.

**Discussion and outlook.** We showed a resonant sensor with the highest TCF reported amongst resonant IR detectors, with a temperature detection sensitivity (63 µK) comparable or better than the performances of the state-of-the-art temperature sensors. We measured an *NETD* as low as 22 mK (in vacuum) and 112 mK (in air) using an optical system having an f-number of 2 (F#=2). It is important to note that we can improve the IR radiation collection efficiency by decreasing the f-number of the IR optics. We could improve the sensitivity 4 times by changing the optics to F#=1. There is room to improve the absorption by a factor of two (e.g. by forming a resonant cavity). With these improvements the NETD would drop to 2.75 mK.

The thermal time constant was measured as $376 \pm 8$ ms in vacuum and $210 \pm 10$ ms in atmospheric pressure, which is slow for video imaging applications. By scaling, one can decrease the thermal time constant (see Supplementary Note 1). Decreasing the radius of the membrane to 52 µm results in a computed time constant of 7 ms, sufficient for video applications (see Supplementary Note 1).

We demonstrated the first use of a phase-change polymer as the transduction mechanism of a highly sensitive IR resonating sensor. The SMP material itself provides the transduction mechanism as well as is a good absorber in the LWIR range (7-14 µm). Our SMP sensors can operate in atmospheric pressure with a very small degradation compared to vacuum operation, due to the high sensitivity of the SMP sensor and to intrinsic high thermal isolation. SMPs can be patterned using established MEMS processes to achieve 50 µm or smaller pixel sizes in large arrays. Capacitive transduction can be used for actuation and for readout. We envision that arrays of SMP resonators can be used as uncooled THz detectors thanks to the high TCF (associated with the material property) and low noise (associated with frequency measurements).

## Methods

**Fabrication.** The fabrication process of the SMP membrane is explained in *Figure 5*. First, a solution of SMP should be prepared to be used in the fabrication. For the devices demonstrated in this work we used MM4520 SMP pellets from SMP Technologies. In order to produce the solution, the SMP pellets are first dissolved in DMF at a weight ratio of 1:5 and mixed overnight at 80°C. The fabrication starts with (1) blade casting of 100nm thick sacrificial Teflon™ AF amorphous fluoroplastic layer *(from Chemours Company)* on an A4-size polyethylene terephthalate (PET) sheet to help the release at the last steps. (2) SMP solution is casted on the sacrificial layer using *Zehntner ZAA2300 film applicator coater*. Each casting step was preceded by an oxygen plasma treatment. After casting, there is a curing step at 80°C on a hotplate for ~3 hours. The process continues with (3) patterning the SMP membrane using a laser cutter (*Trotec Speedy 300*). (4) a PMMA plate with 1 mm thickness was used to form the substrate. (5) An adhesive layer (*ARclear 8932EE, Adhesive Research*) was applied on the PMMA plate. (6) CNC (*CNC Basic 540 from Step-Four*) was used to pattern the PMMA to form circular holes for the resonator. After having the patterned PMMA substrate and the SMP membrane, (7) they were assembled together. (8) PET was cut out to form to suspended SMP resonators. Finally, (9) laser cutter was used to cut out single resonator units.

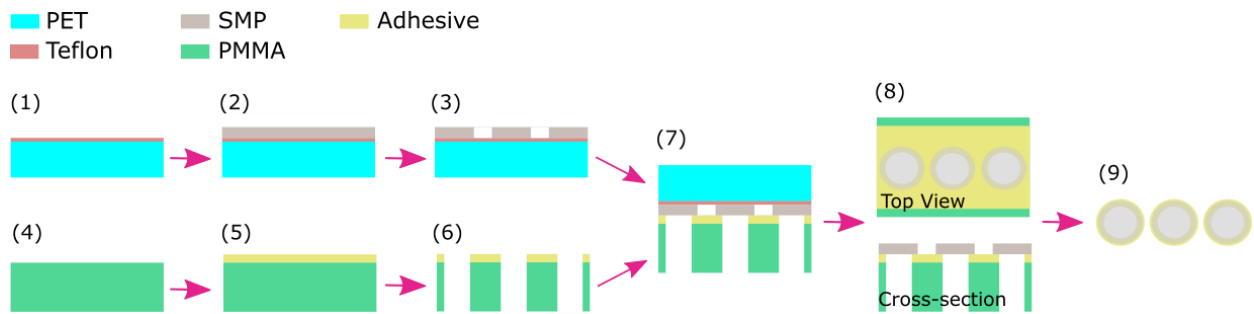

*Figure 5: Schematic illustration for the fabrication process of the SMP resonators. The steps for preparing the solution and casting of the SMP is similar to the one used in ref. 36. The resonator thickness was either* 10 µm or 57 µm.

**Absorber characterization.** Fourier Transform Infrared (FTIR) absorption measurements were performed with THERMO Nicolet 8700 FTIR on bare SMP samples. The spectrum of the transmittance, *T*, was measured. The absorption spectrum, *A*, was then directly extracted from the following expression[45]: *A = 1-T-R* , assuming the total reflectivity (*R*) from the sample in the LWIR is around 5%[46]. .

**Measurement.** Supplementary Figure 4 shows the experimental setup for the IR detection. The SMP resonators were fixed on piezo-disk actuators (*Noliac - NAC 2014, Thorlabs - PA4FEW*), using a thermal adhesive (*Fischer Elektronik - WLFT 404 23X23*). The SMP resonator on the piezo-disk actuator was fixed on a peltier element (*Multicomp - MCPE-127-10-25*) using the same thermal adhesive. A PID temperature controller system (*Meerstetter Engineering GmbH - TEC-1091*) controls and adjusts the operation temperature of the resonators with a PT-100 temperature sensor which was located on the PMMA substrate.

The magnitude and the phase of the frequency response of the SMP resonators were measured using a laser Doppler vibrometer (*Polytec - MSV 400*). The samples were placed into a vacuum chamber, which has a ZnSe optical window without AR-coating (*Crystran LTD - ZNSEP50-3*) to let the visible light for readout and IR radiation for detection. The visible light has normal incidence on the sensor, while the IR radiation has an oblique incidence around 55°- 60° incidence angle. A collimator setup[47] with two AR-coated IR ZnSe lenses (*Thorlabs - LA7656-G - Ø1"- f=25.4 mm and LA7542-G - Ø1"-*

*f=50.1 mm*) were implemented to produce a controllable irradiance that is independent of distance as illustrated in Supplementary Figure 4. Due to space limitations under the LDV, we used the second lens in the collimator with a focal length of 50.1 mm with 25.4 mm diameter that results in a lower collection efficiency due to the use of a lens with f-number of 2. A black painted resistive heater (*DBK - HP04-1/04-24*) was used as an IR source that is placed before a rotating chopper (*Stanford Research Systems, SR540*). The IR target is calibrated with a commercial IR thermometer (PeakTech 4950) to define the temperature difference of the target with the ambient.

The modulated IR radiation from this IR target passes through the IR lens system and illuminates the SMP resonator. This creates a temperature change on the resonator, which changes its Young's modulus and hence the resonance frequency, monitored by the LDV. The internal generator of a lock-in amplifier (*Stanford Research Systems, SR550*) was used to drive the piezo-disk actuator at the fixed driving frequency (around resonance), while the output of the LDV was connected to the same lock-in amplifier to track the phase difference between the actuation signal and the sensor output. Thus, the change in the phase signal is proportional to the temperature difference between the chopper blade and the IR target. In addition, this phase signal was converted to the frequency using the open-loop phase response of the resonator. The slope of the phase response around the resonance frequency ($m = \Delta\varphi/\Delta f$) is determined, then this slope was used to convert the phase to the frequency ($f(t) = \varphi(t)/\mathrm{m}$). At the end, the phase and amplitude data from the lock-in amplifier was collected using an oscilloscope (*Agilent Technologies, MSO9104A*). The Allan Deviation is measured in an open-loop configuration – constant drive frequency. The resonators were not irradiated with IR during the measurements. We followed the same procedure with ref. 22 to calculate the Allan deviation.

## Data availability

The datasets generated during the current study are available from the corresponding author on reasonable request.

**Acknowledgements** We acknowledge financial support from the EPFL. We thank Mario Lehmann for Fourier-transform infrared spectroscopy (FTIR) measurements carried out in the PV-Lab of EPFL.

**Author Contributions** H.S. and J.J.Z. conceived the concept. U.A designed the devices, developed thermal and mechanical models, fabricated the samples, designed and performed the experiments, analyzed and processed the data. T.L. developed the algorithm for the Allan deviation analysis. L.G.V. and H.S. advised on the design of the samples and the experimental methods. U.A wrote the manuscript. All authors read, edited, and discussed the manuscript and agree with the claims made in this work. H.S. coordinated and supervised the research.

## Additional information

**Supplementary Information** accompanies this paper at http://www.nature.com/naturecommunications

**Competing interests:** The authors declare no competing interests.

# Shape Memory Polymer Resonators as Highly Sensitive Uncooled Infrared Detectors

Ulas Adiyan[2], Tom Larsen[2], Juan José Zárate[1], Luis Guillermo Villanueva[2], Herbert Shea[1]

**SUPPLEMENTARY FIGURES**

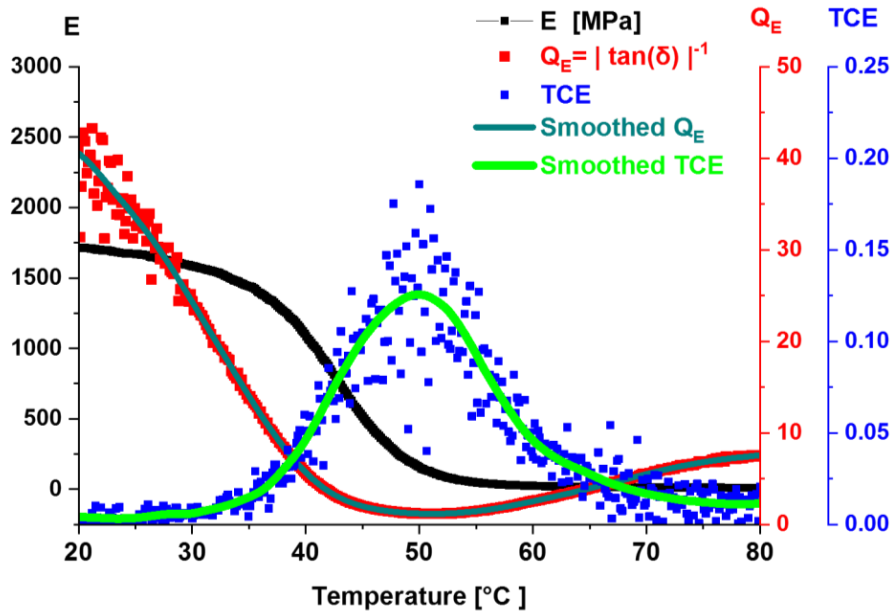

**Supplementary Figure 1:** Dynamic mechanical analysis (DMA) measurement for Young's modulus (E) and intrinsic quality factor ($Q_E$) with respect to temperature for a SMP membrane (*MM4520 SMP pellets from SMP Technologies*). $Q_E$ is the inverse of the loss factor $\tan(\delta)$, $Q_E = |\tan(\delta)|^{-1} = E'/E''$. $E'$ is the storage modulus and $E''$ is the loss modulus. $Q_E$ measurements show low quality factor around the glass transition temperature. The thermal coefficient of the Young's Modulus (TCE) with respect to temperature is plotted in blue, which was calculated from Young's modulus (E) data vs temperature measurement. The smoothed versions of the plots correspond to "50 points" LOESS (locally estimated scatterplot smoothing).

[2] Soft Transducers Laboratory, École Polytechnique Fédérale de Lausanne (EPFL), 2002 Neuchâtel, Switzerland. [2] Advanced NEMS Group, École Polytechnique Fédérale de Lausanne (EPFL), 1015 Lausanne, Switzerland. Correspondence and requests for materials should be addressed to H.S. (email: herbert.shea@epfl.ch)

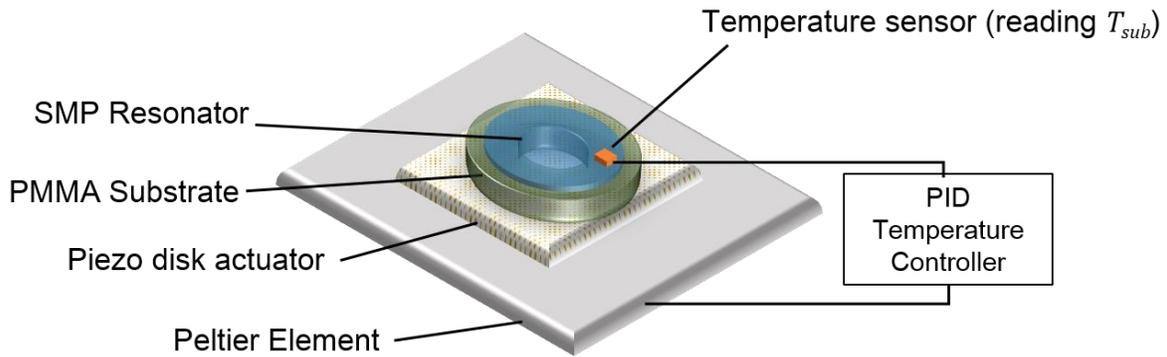

**Supplementary Figure 2:** The SMP resonator (with a radius of 520 µm and a thickness of 10 µm) on a PMMA substrate with a thickness of 1mm. The resonator was attached on a piezo disk actuator and placed on a heater with a PID temperature controller system, which controls the operation temperature of the resonator ($T_{sub}$), using a temperature sensor placed on the substrate.

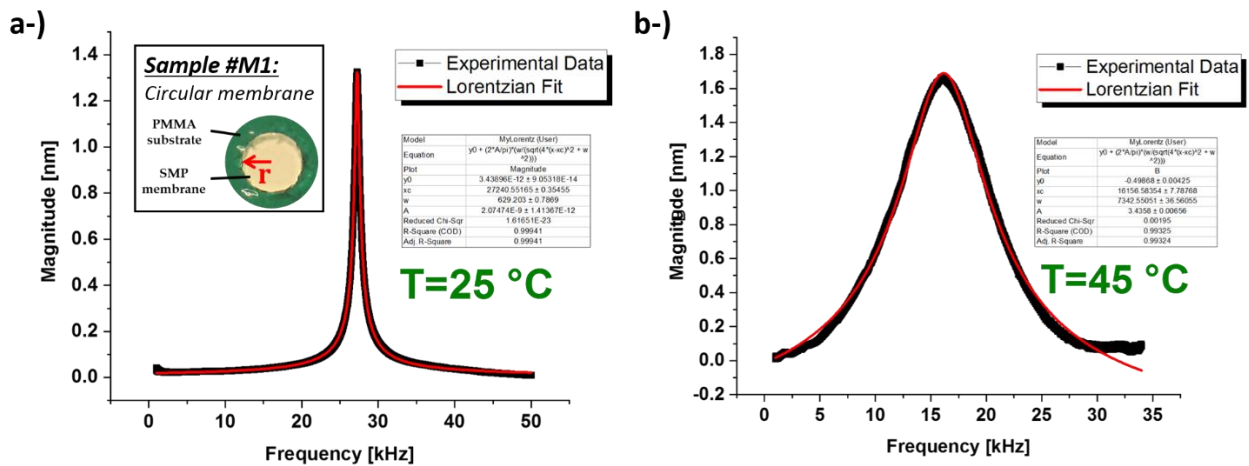

**Supplementary Figure 3:** Displacement vs. frequency of a SMP circular membrane, with radius 520 µm and thickness 10 µm, measured by Laser Doppler Vibrometer, at 25°C and at 45 °C.

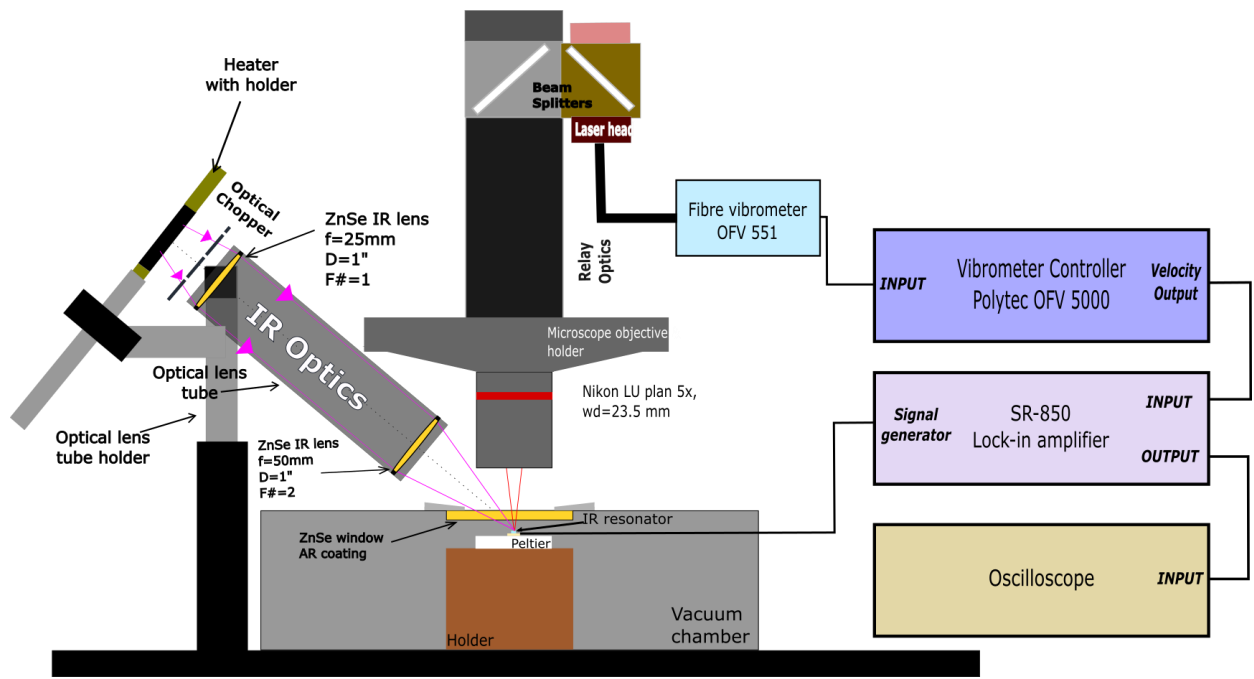

**Supplementary Figure 4:** Schematic of the experimental setup for the characterization of IR SMP resonator sensors to detect the resonant frequency shift of the device subjected to IR radiation.

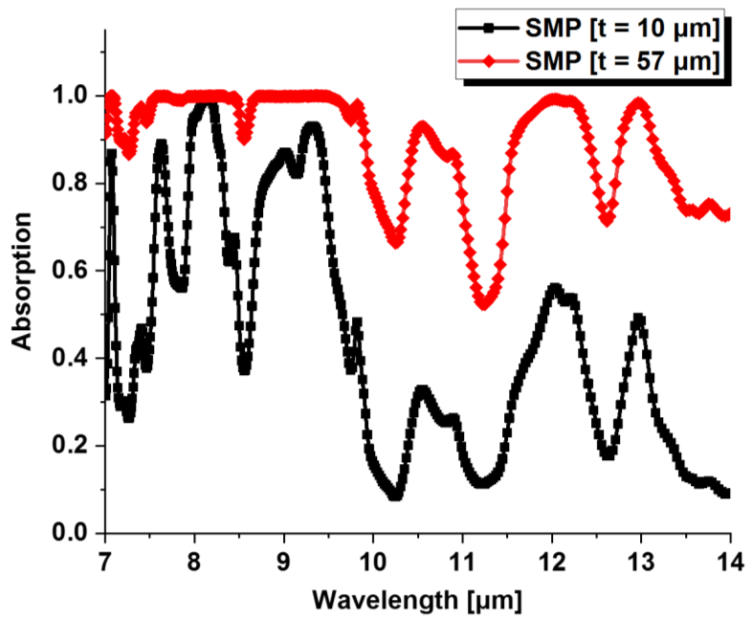

**Supplementary Figure 5:** Absorption spectrum of the 10 μm thick and 57 μm thick SMP sheets in the wavelength range of 7 μm – 14 μm. THERMO Nicolet 8700 FTIR spectrometer was used to obtain the IR absorption of the SMP with different thicknesses.

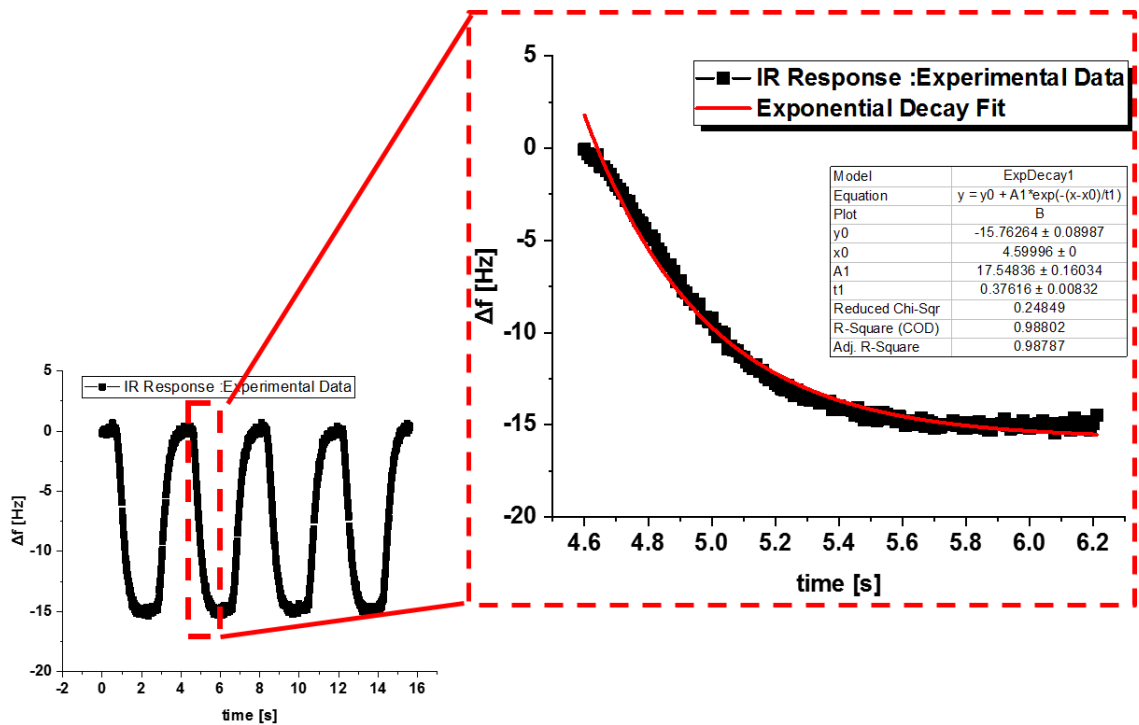

**Supplementary Figure 6:** Thermal time constant of 376 ± 8 ms determined from an exponential fit to the IR response data.

**Supplementary Note 1. Thermal design of the resonator**

A thermal model is essential to estimate the temperature change due to absorbed IR radiation on the sensor as well as to assess the thermal time constant, which determines the speed of the sensor and the integration time for the readout. Supplementary Figure 7 shows the cross-section of the SMP resonator with the developed equivalent circuit model.

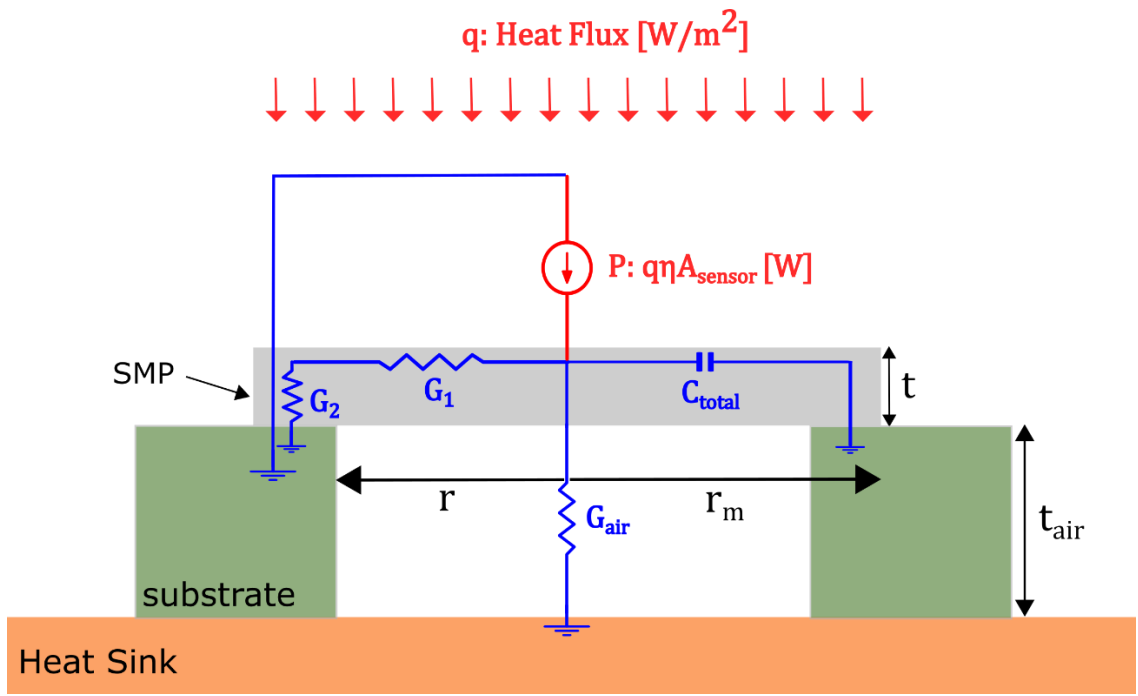

**Supplementary Figure 7:** The thermal model for the SMP resonant IR detector. A heat flux is applied on the membrane and the heat flow from the membrane to the substrate goes through the indicated resistances, which is shown with the equivalent circuit of the thermal model.

Equation (3) shows the temperature change due to the absorbed IR radiation having power of $P$, where $q$ is the heat flux, $\eta$ is the absorption coefficient, $A_{sensor}$ is the sensor absorption area, $G_{total}$ is the total thermal conductance (equivalent thermal conductance), $C_{total}$ is the total thermal capacity and $\omega_{IR}$ is the angular frequency of the modulated IR radiation. As the last part of the Equation (3) implies, for DC or very low modulation frequency ($\omega_{IR} \approx 0$), the temperature change directly depends on the thermal conductance ($G_{total}$), which is generally the case.

$$\Delta T = \frac{P}{\sqrt{G_{total}^2 + \omega_{IR}^2 C_{total}^2}} = \frac{q\eta A_{sensor}}{\sqrt{G_{total}^2 + \omega_{IR}^2 C_{total}^2}} \approx \frac{q\eta A_{sensor}}{G_{total}} \qquad (3)$$

Equation (4) shows the expression for finding the total thermal conductance[1], $G_{total}$. $G_{rad}$ is the radiative thermal conductance, which is originated from the radiative heat exchange between the sensor and the environment. It is the intrinsic thermal conductance of the structure, which emerges as a true limit because all the infrared sensors (cooled/uncooled) are also emitters according to their absorbance based on the Kirchoff's law[2]. $G_1$ and $G_2$ is the thermal conductance through the membrane to the substrate, which is related to the geometry of the structure. The conduction through the air to the substrate is $G_{air}$, which is significant in atmospheric pressure and negligible in vacuum.

$$G_{total} = G_{rad} + \left(\frac{G_1 G_2}{G_1 + G_2}\right) + G_{air} \qquad (4)$$

Equation (5), (6), (7), and (8) provides the thermal conductance for the defined parameters in Supplementary Figure 7 considering a circular membrane. $\sigma$ is the Stefan-Boltzmann constant and $\varepsilon_{SMP}$ is the emissivity of the SMP, which was assumed same as the absorbance[3]. We utilized these equations for our tested circular membrane Sample #M1 to validate the thermal model with the measurement, where the size of the membrane is determined by the ease of the fabrication. Here are the related parameters:

$r$=2.5 mm, $r_m$=520 μm, $t$= 10 μm, $t_{air}$= 1 mm, $k_{SMP}$=0.18 W/(m.K)[4] and $k_{air}$=26.3 x 10$^{-3}$ W/(m.K)[5].

$$G_{rad} = \pi r^2 \sigma \varepsilon_{SMP}(T^4_{ambient} - T^4_{sensor})/(T_{ambient} - T_{sensor}) \quad (5)$$

$$G_1 = k_{SMP}\frac{2\pi rt}{r} = 2\pi k_{SMP} t \quad (6)$$

$$G_2 = k_{SMP}\frac{\pi(r_m^2 - r^2)}{t} \quad (7)$$

$$G_{air} = k_{air}\frac{\pi r^2}{t_{air}} \quad (8)$$

$G_1$ depends only on the thermal conductivity of the SMP and the thickness of the membrane (Equation (6)). For the given geometry, $G_1$ is very small compared to $G_2$, thus, $G_2$ is negligible in the calculations. As it is shown in Supplementary Figure 8.a, $G_{total}$ is significantly determined by $G_1$ in vacuum, $G_1$ and $G_{air}$ in atmospheric pressure. Although $G_{rad}$ is more than 10 times smaller than $G_1$, it has also contribution to the total thermal conductance. For the radius (520 μm) of our sample (Sample M#1), $G_{total}$ is calculated as 1.4 x 10$^{-5}$ W/K in vacuum and 3.6 x 10$^{-5}$ W/K in air. It is very important to note that, the thermal conductance in vacuum and in air is almost same and determined by the thermal conductance of the membrane for smaller radius (<100 μm). It is due to smaller sensor area, which leads to smaller $G_{air}$ and $G_{rad}$. Supplementary Figure 8.b shows the thermal conductance with respect to thickness while keeping the radius fixed (520 μm). Accordingly, the thermal conductance of the membrane and the total thermal conductance (in air and in vacuum) is linearly related with the thickness. Thus, in order to decrease the thermal conductance, one needs to decrease the thickness.

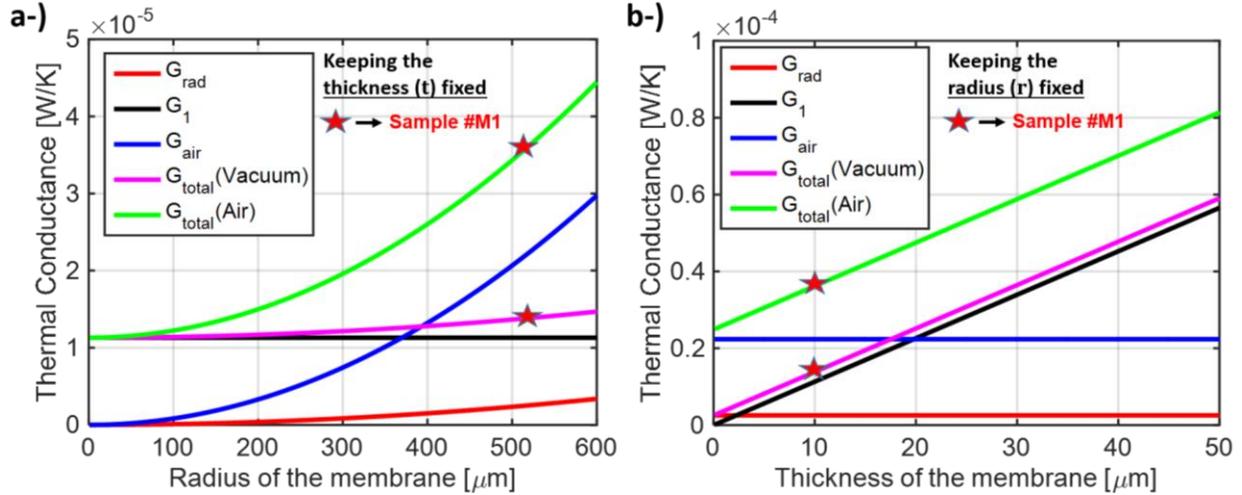

**Supplementary Figure 8**: a-) Thermal conductance vs radius of the membrane of the membrane for a fixed thickness (t= 10 um). b-) Thermal conductance vs thickness of the membrane for a fixed radius (r= 520 um). The contribution of the conduction through radiation, through air, through the membrane and the total thermal conductance in vacuum and in air were plotted for the operation temperature T = 30 °C. The red star shows the tested sample #M1 on both plots.

Following the analysis of the total thermal conductance, the total heat capacity of the membrane should also be determined in order to find the thermal time constant ($\tau_{th}$) of the membrane. The calculation of the total heat capacity of the membrane is expressed in Equation (9), where $\rho$ refers to the density of the SMP membrane which was measured as 1100 kg/m³. c refers to the specific heat capacity, and has values changing from 1200 J/(kg.K) to 1800 J/(kg.K) between 25 °C to 50 °C according to the differential scanning calorimetry (DSC) measurements of SMP[6].

$$C_{total} = (\pi r^2 t)\rho\, c \tag{9}$$

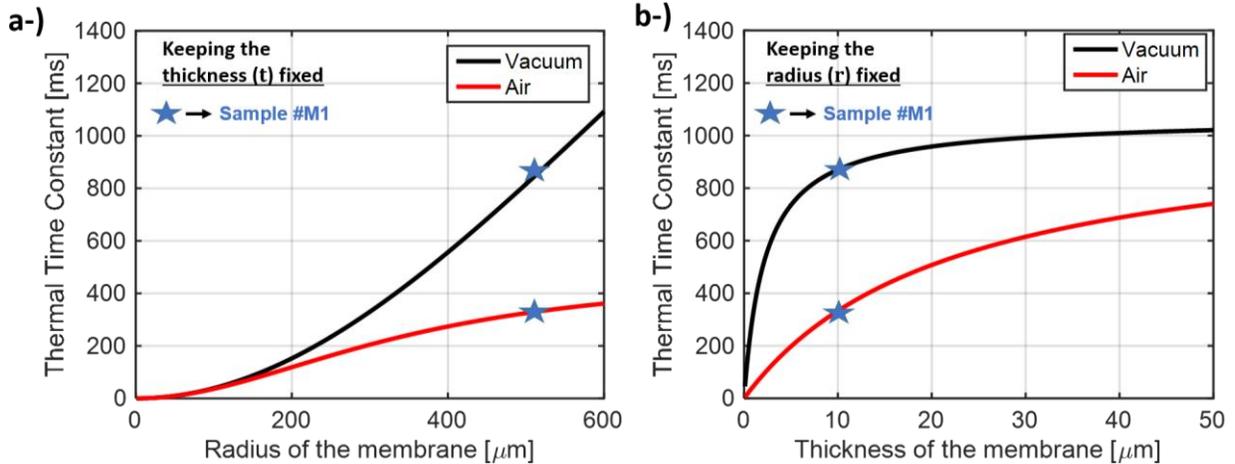

**Supplementary Figure 9**: a-) Thermal time constant vs radius of the membrane of the membrane for a fixed thickness (t= 10 um). b-) Thermal time constant vs thickness of the membrane for a fixed radius (r= 520 um). The thermal time constant in vacuum and in air were plotted for the operation temperature T = 30 °C. The blue star shows the tested sample, Sample #M1 on both plots.

The thermal time constant of the SMP membrane, can be expressed as $\tau_{th} = \frac{C_{total}}{G_{total}}$. Supplementary Figure 9.a and Supplementary Figure 9.b analyzes the effect of the radius and the thickness of the membrane on the thermal time constant respectively. In order to decrease the time constant, one needs to decrease the radius and/or the thickness of the membrane, while considering the effect of both parameters on NETD via the thermal conductance and the absorber area.

As the star indicators spot out in Supplementary Figure 9, thermal time constant are 335 ms in air and 885 ms in vacuum (at T=30 °C) from the thermal model, where the exponential fit from the measurements reveal time constants of 210 $\pm$ 10 ms and 376 $\pm$ 8 ms. The differences in time constants between the experimental data and the thermal model analysis can be attributed to the differences in material properties of the SMP in real life and in the calculations. Although, these time constant values seem high for video applications, it is straightforward to decrease the time constant. For instance, 10 times decrease in the radius (from 520 µm to 52 µm) makes the thermal time constant 100 times lower in vacuum, results in 7 ms according to our model, which is quite sufficient for video applications. In addition, according to our model, this reduction results in almost same time constant in air. This is very important because the thermal conductance through air becomes negligible (very small compared to thermal conductance through the membrane) for these dimensions; which results in the same high performance for atmospheric pressure condition as it is in vacuum (Supplementary Figure 8.a). The reason we preferred larger membranes is completely related to the present fabrication capabilities (See Methods for the fabrication).

Consequently, it is fairly straightforward to adjust the speed of the sensor for video applications and enable the operation of the sensor in atmospheric pressure by maintaining the same performance in vacuum. Decreasing the radius of the membrane will decrease the absorption area and degrade the performance. Therefore, some modifications for the dimensions are essential such as decreasing the thickness of the SMP membrane, which results in a decrease in thermal conductance and balances the degradation the performance while still evaluating the thermal time constant to be compatible with video applications.

**Supplementary Note 2 for NETD calculation**

The noise-equivalent temperature difference (NETD) is the most widely used figure of merit for the performance of IR sensor systems. It determines the minimum detectable temperature difference for an IR target (i.e., IR object, the radiation source). In order to find the NETD[7], one divides the temperature difference of the IR target with the ambient by the SNR of the detection system to obtain the minimum detectable temperature difference for an SNR = 1. For resonant IR sensors, the TCF determines the signal level for the SNR calculations. The overall noise of the system is the second parameter that defines the SNR, which can be extracted from Allan deviation ($\sigma_A$) measurements. Therefore, $\frac{\sigma_A}{TCF}$ gives the minimum detectable temperature for an SNR of 1, which corresponds to the direct temperature sensing as a temperature sensor.

In order to convert this minimum detectable temperature to the IR target (i.e. IR source or object) plane, one needs to consider the temperature change on the sensor that corresponds to the change on the IR target. There are many different parameters that influence the detector to target temperature ratio such as the transmission of the IR optical system (includes atmospheric transmission, transmission of the optical windows and lenses), the collection efficiency of the IR lens system and emissivity of the IR target ($\varepsilon = 1$, assuming an ideal blackbody), the absorbance and area of the IR sensor. Equation (10) represents the detector to target temperature ratio ($\frac{\Delta T_D}{\Delta T_T}$), where $\eta$ is the absorbance of the detector, $\tau_0$ is the transmittance of the IR optical system through atmosphere and IR optics), $(dP/dT)_{\lambda 1-\lambda 2}$ is the radiated power change per temperature change for the IR wavelengths between $\lambda 1$=8 µm and $\lambda 2$=14 µm per unit area [SI unit: W

m$^{-2}$K$^{-1}$], $F_{\#}$ is the f-number of the IR lens system and $G_{total}$ [SI unit: W m$^{-1}$] is the total thermal conductance of the system[8]:

$$\frac{\Delta T_D}{\Delta T_T} \approx \frac{\eta \tau_0 A_{sensor}(dP/dT)_{\lambda 1-\lambda 2}}{4F_{\#}^2 G_{total}} \qquad (10)$$

After using Equation (10) for our tested sample Sample #M1, we found the $\Delta T_D/\Delta T_T$ as ~ 0.003, which shows that 1 °C temperature change on the IR target causes 3.3 m°C change on the sensor in vacuum. Similarly, 1 °C temperature change on the IR target causes 1.2 m°C in air. In this calculation, $\tau_0$ is 0.7 according to the specifications of the optical window and lenses, $\eta$ is 0.44 according to FTIR measurement, $(dP/dT)_{\lambda 1-\lambda 2}$ is 2.62 Wm$^{-2}$K$^{-1}$ for the operation wavelength range[8], $F_{\#}$ is 2 and $G_{total}$ was already calculated for vacuum and air.

By optimizing the parameters such as thermal conductance and absorbance, and by using an IR optical system with higher collection efficiency (with low $F_{\#}$), one can increase this detector to target temperature ratio so the sensitivity of the IR sensor. In order to define the sensitivity of the system one needs to find the overall noise of the system. To this aim, many different sources that may contribute to the total noise of the resonant IR detectors should be considered as they are discussed in the following parts.

**Background Fluctuation Noise**

There is a fundamental limit in the thermal isolation of the sensor that is caused by the radiative heat exchange. The temperature change of the background leads to continuing heat exchange between the sensor and its environment through radiation. It is important to express that, all type of IR sensors aim to absorb IR radiation, which also makes them good emitters for the same waveband[3]. Thus, any IR detector (cooled or uncooled) cannot avoid from this radiative heat exchange, which sets the fundamental limit for all IR detectors. Equation (11) shows the $NETD_{BF}$, which detector to target temperature ratio was already included to the expression[9].

$$NETD_{BF} = \frac{8F_\#^2}{\eta \tau_0 (dP/dT)_{\lambda 1 - \lambda 2}} \sqrt{\frac{2k_B \sigma_T B (T_D^5 + T_B^5)}{A_{sensor}}} \qquad (11)$$

Some of the parameters in the expression were already defined. Other than these parameters, the $NETD_{BF}$ depends on the $k_B$ (Boltzmann's constant - [SI unit: JK$^{-1}$]), $\sigma_T$ (Stephan-Boltzmann constant - [SI unit: Wm$^{-2}$K$^{-4}$]), $B$ (measurement bandwidth [SI unit: Hz]), $T_D$ and $T_B$ (Detector and background temperature respectively [SI unit: K]).

**Temperature Fluctuation Noise**

In addition to the temporal fluctuations of temperature due to the radiative heat exchange, the thermal conduction through the membrane or the air to the substrate constitutes as another fundamental limitation for all thermal detectors (only uncooled detectors). In order to eliminate the effect of the thermal conductance through the air, most of the thermal detectors typically operate at low pressures (~1e-5 mbar). Since there are temporal fluctuations of temperature at the membrane by means of thermal conduction through the membrane to the substrate, it is called as the temperature fluctuation noise. Equation (12) shows the expression of the $NETD_{TF}$ related to the temperature fluctuation noise. All of the parameters used in Equation (12) were also indicated in the previous subsections.

$$NETD_{TF} = \frac{8F_\#^2 T_D \sqrt{k_B B G_{total}}}{\eta \tau_0 A_d (dP/dT)_{\lambda1-\lambda2}} \tag{12}$$

**Thermomechanical noise**

Thermomechanical noise is another fundamental noise source for mechanically movable thermal detectors apart from background and temperature fluctuation noise. The vibrational noise originated from the thermal energy ($k_B T$) causes the thermomechanical noise through the continuous exchange of the mechanical energy on the sensor. Equation (13) shows the expression[10] for the $NETD_{TM}$, where $E_c$ refers to the carrier energy representing the maximum drive energy[11].

$$NETD_{TM} = \frac{4F_\#^2 G}{\eta \tau_0 A_d (dP/dT)_{\lambda1-\lambda2}} \frac{1}{TCF w_0} \sqrt{\frac{k_B T B w_0}{E_c Q}} \tag{13}$$

$$E_c = M_{eff} w_0^2 \langle x_c^2 \rangle \tag{14}$$

Equation (14) shows the expression of the carrier energy for a circular membrane, where $M_{eff}$ is the effective mass of the membrane (0.2695m for the flexural mode (0,1) for a circular membrane[12], where m is the mass of the membrane) and $x_c$ is the maximum displacement can be achieved by our circular membranes ($x_c$ is accepted as 100 nm).

For the calculation of the $NETD_{TM}$ for different thicknesses (at T=30 °C), $w_0$ was used from the measurements ($w_0 = 2\pi f_{Res}$, where $f_{Res}$ was measured as 25.53 kHz for our membrane having a thickness of 10 um) and then $w_0$ was modified linearly for the other thicknesses. The Q-factor and TCF obtained from the measurements are Q=34 and 1%/°C respectively and they were accepted same for all thicknesses. The measurement bandwidth ($B$) is taken as 1 Hz to be coherent with the thermal time constant of Sample M#1. The other parameters were already introduced in the calculation of detector to target temperature ratio ($\frac{\Delta T_D}{\Delta T_T}$).

The explained fundamental noise sources contribute to the total noise of the system independently and determine the sensitivity of our resonant IR sensors. All these noise sources are independent random processes, hence the root mean square value of the noise sources were calculated as the NETD for the whole system as $NETD_{TOTAL}$ in Equation (15).

$$NETD_{TOTAL} = \sqrt{NETD_{BF}^2 + NETD_{TF}^2 + NETD_{TM}^2} \tag{15}$$

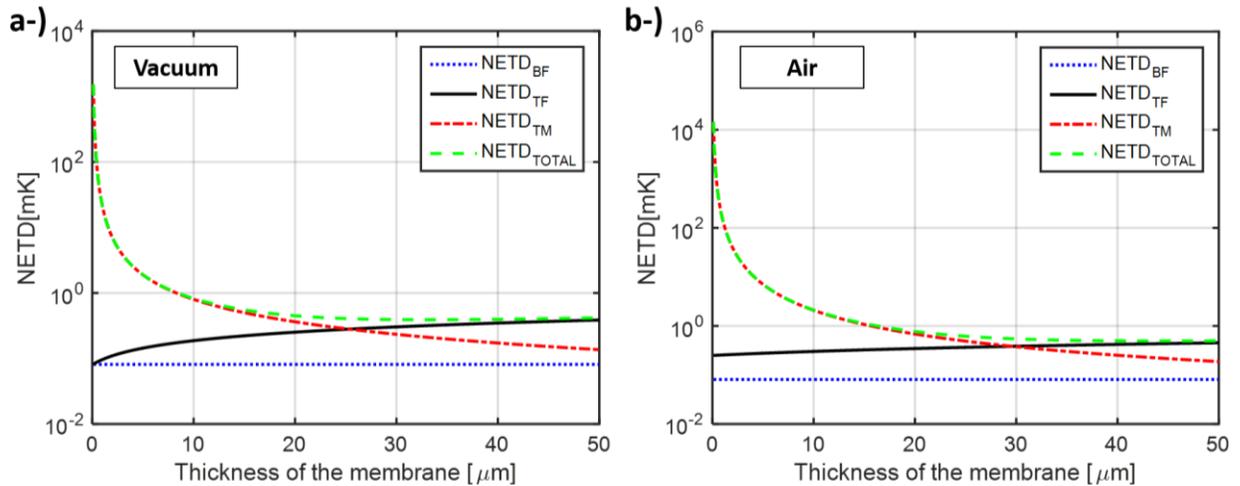

**Supplementary Figure 10:** NETD calculation vs. the thickness of the membrane for a fixed radius (r= 520 um – which is the radius of Sample #M1) at the operation temperature T = 30 °C. a-) in vacuum b-) in air. $NETD_{BF}$: Background fluctuation noise, $NETD_{TF}$: Thermal fluctuation noise, $NETD_{TM}$: Thermomechanical noise.

Supplementary Figure 10 shows our calculations for the fundamental NETD noise sources with the total NETD with respect to thickness of the membrane, while keeping the radius of the membrane fixed at 520 µm. Considering the Sample #M1 in vacuum, the calculated total NETD is limited by the thermomechanical (TM) noise to a certain thickness value (~ 20 µm). The reason for the huge TM noise for small thicknesses (<2 µm) are due to the decrease in the resonance frequency with thickness. On the other hand, in atmospheric pressure, the NETD performance, for the thicknesses smaller than 30 µm, slightly degrades as compared to vacuum due to thermal conductance of air.

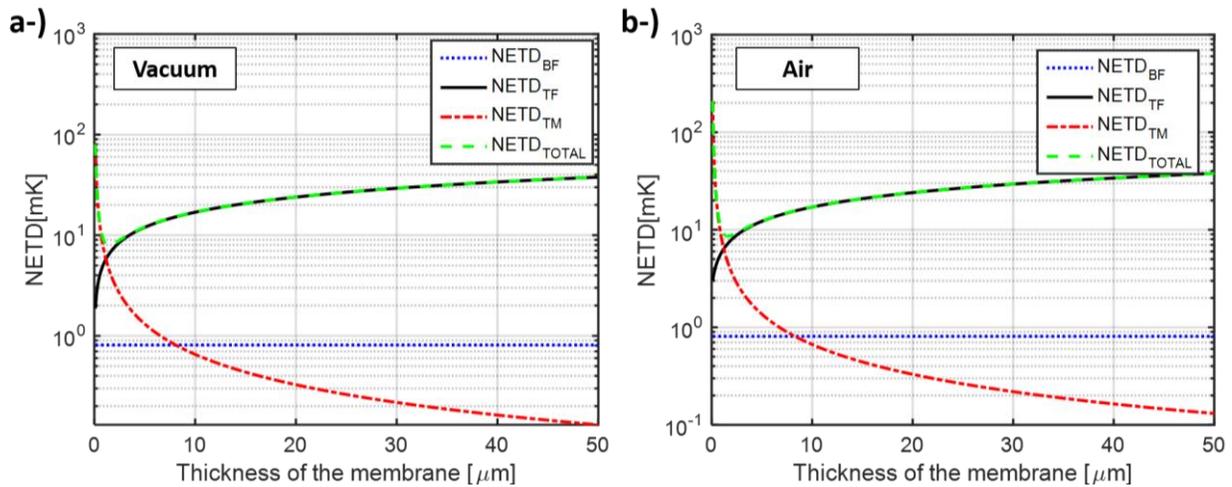

**Supplementary Figure 11:** NETD calculation vs the thickness of the membrane for a fixed radius (r= 52 um – which refers to 10 times reduction for supporting real time applications) at the operation temperature T = 30 °C. a-) in vacuum b-) in air. $NETD_{BF}$: Background fluctuation noise, $NETD_{TF}$: Thermal fluctuation noise, $NETD_{TM}$: Thermomechanical noise.

In the Supplementary Note 1 on the thermal design, we explained how the speed of the sensor can be adjusted for video applications by decreasing the radius of the membrane. In order to evaluate the effect of the thickness on the performance of the IR sensor, the fundamental NETD noise sources with the total NETD were analyzed in regard to the thickness of the membrane (Supplementary Figure 11). In addition, in the case of the radius reduction to 52 µm, the NETD calculations were plotted in vacuum and in air to determine the thickness. As Supplementary Figure 11 shows, the optimum thickness for a membrane with a radius of 52 μm, is around 1.5 μm; which would provide 9 mK NETD for F#2 and 50% absorption in vacuum as well as in air, despite the reduction in the absorber area around 100 times.

All these noise sources have an effect on the frequency stability of our IR resonant sensors. But the thermomechanical noise and the temperature fluctuation noise are the most significant ones for our system, which set a fundamental noise limit for these sensors independent from the read-out noise. Furthermore, the radius reduction at the membrane to achieve high speed for video applications should be accompanied with thickness reduction to attain high performance.

**Supplementary Note 3 for NETD measurements**

The measured NETDs in Figure 4.c of the manuscripts can be compared with a predicted NETD, which is the combination of the measured temperature detection sensitivity with the theoretical value of $\frac{\Delta T_D}{\Delta T_T}$, the detector to target temperature ratio (see Supplementary Note 2). In order to calculate the predicted NETD, Equation (16) was utilized.

$$NETD = \frac{\Delta T_D}{\Delta T_T} \frac{\sigma_A}{TCF} \qquad (16)$$

Thus, for the predicted NETD, we used the $\frac{\sigma_A}{TCF}$ measurements from Figure3.c of the manuscript and the computed theoretical $\frac{\Delta T_D}{\Delta T_T}$ using the specifications and assumptions based on the IR optical system.

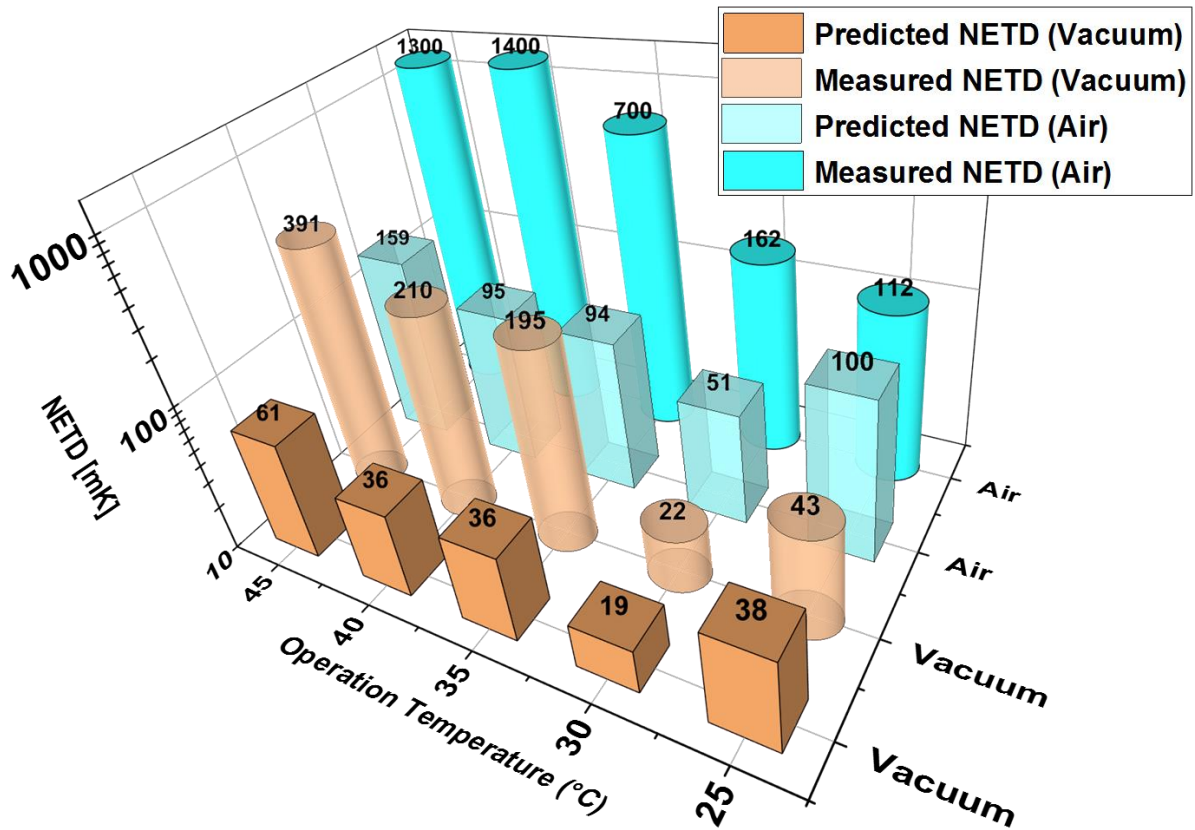

**Supplementary Figure 12:** The measured NETD vs operation temperature in vacuum and in air, along with the predicted NETD.

The best measured NETD is 22 mK (in vacuum, at *30 °C*) and 112 mK (in air, *25 °C*) using an optical system having an f-number of 2 (F#=2). Although these values are consistent with the predicted NETD, the difference between the measured and the predicted NETD increases as the operation temperature increases (Supplementary Figure 12). In our predicted NETD calculations we assumed a fixed absorption and a fixed thermal conductivity for the SMP. These values may be temperature dependent, which could explain part of the difference.